\documentclass[prd,aps,nofootinbib,superscriptaddress,floatfix,preprintnumbers,twocolumn]{revtex4}

\usepackage[colorlinks, citecolor=blue,anchorcolor=blue,menucolor=blue, linkcolor=blue,filecolor=blue,runcolor=blue,urlcolor=blue,frenchlinks=true]{hyperref}
\usepackage{amsmath,amsfonts,amssymb,mathtools,cases,slashed,bm}
\usepackage{epsfig, graphicx}
\usepackage{color}
\usepackage[dvipsnames]{xcolor}
\usepackage{tabularx}
\usepackage{physics}
\usepackage{hhline}
\usepackage{mathrsfs}
\usepackage[normalem]{ulem}
\usepackage{multirow}
\usepackage{float}
\usepackage{appendix}
\usepackage{orcidlink}
\usepackage[normalem]{ulem}

\usepackage{verbatim}

\begin{document}

\title{First Self-Renormalized Gluon PDF of Nucleon from Large-Momentum Effective Theory in the Continuum Limit}

\author{Alex NieMiera \orcidlink{0009-0007-0553-2590}}
\email{niemiera@msu.edu}
\affiliation{Department of Physics and Astronomy, Michigan State University, East Lansing, Michigan 48824, USA}
\affiliation{Department of Computational Mathematics,
  Science and Engineering, Michigan State University, East Lansing, Michigan 48824, USA}

\author{William Good \orcidlink{0000-0001-8428-1003}}
\email{goodwil9@msu.edu}
\affiliation{Department of Physics and Astronomy, Michigan State University, East Lansing, Michigan 48824, USA}
\affiliation{Department of Computational Mathematics,
  Science and Engineering, Michigan State University, East Lansing, Michigan 48824, USA}

\author{Huey-Wen Lin \orcidlink{0000-0001-6281-944X}}
\affiliation{Department of Physics and Astronomy, Michigan State University, East Lansing, Michigan 48824, USA}

\author{Fei Yao \orcidlink{0000-0002-7227-0996}}
\affiliation{Physics Department, Brookhaven National Laboratory, Upton, New York 11973, USA}


\begin{abstract}
We present the first lattice-QCD determination of the nucleon gluon parton distribution function (PDF) within the large-momentum effective theory (LaMET) framework, employing the hybrid scheme with self renormalization, in the continuum limit.
High statistics calculations with boost momentum $P_z \approx 2.0$-$2.2$~GeV are performed at lattice spacings of $a \approx \{ 0.15, 0.12, 0.09 \}$~fm, a pion mass of $M_\pi \approx 310$~MeV, with $N_f=2+1+1$ quark flavors. 
We apply gradient flow smearing with $\mathcal{T}_\text{W} = 3a^2$, finding that self-renormalization remains stable.
The renormalized gluon matrix elements and PDFs show consistent behavior across lattice spacings, with the $a \approx 0.09$~fm results statistically compatible with the continuum limit.
Our final LaMET gluon PDF compares reasonably with select global-fit determinations, specifically preferring analyses which have near-zero gluonic density in the $x \gtrsim 0.6$ region.
\end{abstract}
\preprint{MSUHEP-25-024} 
\date{\today}
\maketitle

\textit{Introduction. ---}
The gluon parton distribution function (PDF) is a cornerstone of modern quantum chromodynamics (QCD), describing the momentum distribution of gluons within hadrons and nuclei.
Precise knowledge of gluon PDFs is critical for reliable predictions of hard-scattering processes at colliders, including Higgs-boson production, jet formation, and heavy-flavor dynamics.
Since gluons dominate at small Bjorken-$x$, uncertainties in their distributions directly limit the accuracy of Standard-Model cross sections and the sensitivity of new-physics searches.
In the large-$x$ region, gluons play a critical role in processes that probe high momentum transfer and heavy mass scales, such as top-quark pair production and searches for heavy resonances, and uncertainties in their PDFs directly propagate into cross-section predictions and theoretical error budgets for the LHC and future colliders.
Despite this importance, the large-$x$ gluon PDF remains among the least constrained aspects of proton structure, mainly due to the scarcity of high-precision measurements that directly probe gluon dynamics at large $x$~\cite{Achenbach:2023pba,Amoroso:2022eow}.
Ongoing and future experimental programs aim to address these knowledge gaps:
the LHC continues to provide constraints across small and large $x$, fixed-target experiments cover the intermediate-$x$ region, and the upcoming Electron-Ion Collider (EIC) will enable precise mapping of gluon distributions in protons and nuclei~\cite{AbdulKhalek:2021gbh,Accardi:2012qut,Burkert:2022hjz}.

Lattice QCD provides a first-principles, nonperturbative framework for studying PDFs by discretizing QCD on a spacetime lattice and performing numerical simulations.
While traditional global fits rely on experimental cross sections and perturbative QCD factorization, lattice QCD provides a complementary \textit{ab initio} approach, especially valuable in regions with scarce or uncertain constraints, such as the large-$x$ region ($x \gtrsim 0.4$) relevant for new physics searches and high transverse momentum predictions.
Lattice-QCD techniques, such as large-momentum effective theory (LaMET)~\cite{Ji:2013dva} and pseudo-PDF~\cite{Balitsky:2019krf} approaches, enable the extraction of moments and $x$-dependent distributions of many isovector nucleon and valence pion quark PDFs and GPDs;
We refer readers interested in learning more about these topics to the reviews in Refs.~\cite{Ji:2020ect,Constantinou:2020hdm,Lin:2023kxn,Lin:2025hka}.
In contrast, fewer gluon PDF calculations exist with progress mainly limited to the pseudo-PDF approach: nucleon~\cite{Fan:2020cpa, Fan:2022kcb, HadStruc:2021wmh, Delmar:2023agv, HadStruc:2022yaw}, pion~\cite{Good:2023ecp}, and kaon~\cite{Salas-Chavira:2021wui,NieMiera:2025inn} gluon PDFs.
Only recently have the first determinations of the nucleon gluon PDFs been achieved using LaMET~\cite{Good:2024iur,Good:2025daz}.

In this paper, we report the first calculation of the nucleon gluon PDF from LaMET using matrix elements renormalized in the hybrid scheme~\cite{Ji:2020brr} with self renormalization~\cite{LatticePartonCollaborationLPC:2021xdx}.
This methodology is regarded as one of the most robust ways to renormalize the lattice matrix elements, handling several sources of divergences and minimizing discretization effects.
Previous lattice studies using self-renormalization have focused primarily on quark PDFs, including transversity~\cite{LatticeParton:2022xsd}, helicity~\cite{Holligan:2024wpv}, and pion and kaon distribution amplitudes~\cite{Holligan:2023rex,LatticeParton:2022zqc}.
In this work, we further advance determinations of gluon PDFs from LaMET by applying the self-renormalization method across multiple lattice spacings.

\textit{Lattice details. ---}
We use high-statistics data on three ensembles with lattice spacings $a \approx \{ 0.15, 0.12, 0.09 \}$~fm at a valence pion mass $M_{\pi} \approx 310$~MeV generated by the MILC collaboration~\cite{MILC:2013znn} using $N_f = 2+1+1$ highly improved staggered quarks (HISQ)~\cite{Follana:2007rc}.
We use Wilson clover fermions in the valence sector and tune the valence-quark mass to reproduce the light and strange masses in the HISQ sea.
We use the same mixed-action setup as PNDME Collaboration, which has been applied to precision nucleon-structure studies across up to five lattice spacings and multiple physical pion mass ensembles~\cite{Park:2025rxi,Jang:2023zts,Mondal:2020cmt,Jang:2019jkn,Gupta:2018lvp,Lin:2018obj,Gupta:2018qil,Bhattacharya:2015wna,Bhattacharya:2015esa,Bhattacharya:2013ehc}, and we observe no exceptional-configuration effects in any two- or three-point correlators across the three ensembles used in our study.

We computed $O(10^5)$ two-point correlators for $\eta_s$ meson, with $M_{\eta_s} \approx 690$~MeV at zero momentum to calculate the self-renormalization factors, and measured $O(10^6)$ nucleon two-point correlators, with Gaussian momentum smearing~\cite{Bali:2016lva} to obtain nucleon matrix elements with boost momenta $P_z \approx 2.0-2.2$~GeV.
The interpolating operators we use are $\overline{s}\gamma_t\gamma_5 s$ for $\eta_s$ and $\epsilon_{abc}(d^T_aC \gamma_5u_b)\mathcal{P}_+ u_c$ for the nucleon, where $C$ denotes the charge-conjugation operator and $\mathcal{P}_+$ is the positive-parity projector.
Alternative operators with improved performance at high momentum were proposed in Ref.~\cite{Zhang:2025hyo}; however, much of the data used in this work predates their publication, and the signal enhancement is expected to be smaller for heavier than physical pion mass calculations, such as this.
Further lattice details for this study are summarized in Table~\ref{tab:lattice_details}.
Following previous work by MSULat~\cite{Fan:2020cpa, Fan:2022kcb, Good:2023ecp, Salas-Chavira:2021wui,NieMiera:2025inn,Good:2024iur,Good:2025daz}, we extract matrix elements from two- and three-point correlators using two-state simultaneous fits and tune the fit ranges to minimize excited state contamination.

Furthermore, reasonable signal for the gluon correlators has not been obtained without the employment of gauge-link smearing~\cite{Fan:2020cpa, Fan:2022kcb, HadStruc:2021wmh, Delmar:2023agv, HadStruc:2022yaw, Good:2023ecp, Salas-Chavira:2021wui,NieMiera:2025inn,Good:2024iur,Good:2025daz}.
However, self-renormalization has only been analytically proven for zero smearing and shown to work reasonably well with a small amount of smearing ~\cite{LatticePartonLPC:2021gpi}.
In this work, we utilize Wilson flow~\cite{Luscher:2010iy} with a fixed relative flow time of $\mathcal{T}_\text{W} = 3 a^2$.
Applying Wilson flow changes the relevant scale of the problem from $1/a$ to $1/\sqrt{8\mathcal{T}_\text{W}} \propto 1/a$~\cite{Luscher:2010iy}, hence there is motivation that self-renormalization should work on this setup without modification.
Additionally, because $a \rightarrow 0$ naturally drives $\mathcal{T}_\text{W} \rightarrow 0$ in this setup, one would naively expect that taking the continuum limit doubles as a zero-flow time extrapolation.
Commutativity of these limits should be studied more carefully by approaching the physical point via different trajectories, as will be explored in forthcoming work~\cite{NieMiera:2025xxx}.

\begin{table}[t]
    \centering
    \renewcommand{\arraystretch}{1.3}
    \begin{tabular}{| c | c | c | c |}
    \hline
        Ensemble & a$09$m$310$ & a$12$m$310$ & a$15$m$310$ \\ \hline
        $a$ (fm) & $0.0888(8)$ &  $0.1207(11)$ & $0.1510(20)$ \\ \hline
        $L^3 \cross T$ & $32^3 \cross 96$ & $24^3 \cross 64$ & $16^3 \cross 48$ \\ \hline
        $M_\pi^\text{val}$ (GeV) & $0.313(1)$ & $0.309(1)$ & $0.319(3)$ \\ \hline
        $M_{\eta_s}^\text{val}$ (GeV) & $0.698(7)$ & $0.6841(6)$ & $0.687(1)$ \\ \hline
        $N_\text{cfg}$ & $1026$ & $1013$ & $900$ \\ \hline
        $N_\text{meas}^\text{2pt}$ ($P_z = 0$) & $196,992$ & $64,832$ & $129,600$ \\ \hline
        $N_\text{meas}^\text{2pt}$ ($P_z \neq 0$) & $1,378,944$ & $1,555,968$ & $950,400$ \\ \hline
    \end{tabular}
    \caption{Lattice spacing ($a$), lattice volume ($L^3\times T$), valence pion mass ($M_\pi^\text{val}$), $\eta_s$ mass ($M_{\eta_s}^\text{val}$), number of configurations $N_\text{cfg}$, and two-point correlators measurements for zero momentum ($N_\text{meas}^\text{2pt}$ ($P_z = 0$)) and nonzero momentum ($N_\text{meas}^\text{2pt}$ ($P_z \neq 0$)) data for each HISQ ensemble generated by the MILC Collaboration and analyzed in this study.}
    \label{tab:lattice_details}
\end{table}

\textit{Self-renormalization factor. ---}
In LaMET, the gluon PDF is obtained by first measuring the bare matrix elements $\tilde{h}^B(z, P_z, 1/a) = \langle P_z | O(z) | P_z \rangle$ on the lattice, where $|P_z\rangle$ is a hadron state with boost momentum $P_z$, and $O(z)$ is an operator correlating gluon fields across a Euclidean separation $z$.
Several gluon operators exist which are multiplicatively renormalizable and go to the same lightcone limit~\cite{Zhang:2018diq,Balitsky:2019krf}, but we have found in previous work~\cite{Good:2024iur} that the following operator performs significantly better than others:
\begin{equation} \label{eq:gluon-operator}
    O (z) =
    F^{ti}(z)U(z,0)F_i^t(0) - F^{ij}(z) U(z,0) F_{ij}(0),
\end{equation}
where $i,j$ denote the summation over transverse indices $\{x,y\}$, the Wilson line $U(z,0)$ ensures gauge invariance, and $F_\text{a}^{\mu \alpha} = \partial^\mu A_\text{a}^\alpha - \partial^\alpha A_\text{a}^\mu - g f_\text{abc} A_b^\mu A_\text{c}^\alpha$ is the gluon field strength tensor.

To remove the ultraviolet (UV) divergences in the bare quasi correlation $\tilde{h}^B(z, P_z, 1/a)$, a multiplicative renormalization may be applied~\cite{Zhang:2018diq}
\begin{equation} \label{eq:multiplicative-renormalization}
    \tilde{h}^B(z, P_z, 1/a)
    =
    Z_L(a) e^{- \delta m(a) z} \tilde{h}^R (z, P_z, 1/a),
\end{equation}
where $\tilde{h}^R (z, P_z, 1/a)$ is the renormalized quasi correlation. 
Here, $Z_L(a)$ removes the logarithmic divergences that are independent of the Wilson-line length $z$ while $\delta m(a)$ cancels the linear divergences proportional to $z$ arising from the Wilson-line self-energy.
Purely nonperturbative renormalization schemes all suffer from the problem that the renormalization factor introduces unwanted nonperturbative contributions, distorting the infrared behavior of the original quasi-correlation.
However, the hybrid renormalization scheme~\cite{Ji:2020brr} addresses this issue by handling the short and long distance contributions separately, with the renormalized correlation expressed as
\begin{multline} \label{eq: renormalized-quasi-LF}
    \tilde{h}^R (z, P_z)
    =
    \frac{\tilde{h}^B(z, P_z, 1/a)}{\tilde{h}^B(z, P_z = 0, 1/a)} \theta (z_s - |z|) + \\ T_s  \frac{\tilde{h}^B (z, P_z, 1/a)}{Z_R (z, 1/a)} \theta (|z| - z_s),
\end{multline}
where $z_s$ defines the boundary between short- and long-distance regions, and $T_s$ is a scheme conversion factor to ensure continuity at $z = z_s$.
For short distances ($z < z_s$), we employ the ratio method~\cite{Radyushkin:2018cvn} to remove the $z$-dependent UV divergences associated with the Wilson line.
For larger separations ($z > z_s$), self-renormalization is applied.
In this approach, the self-renormalization factor $Z_R(z, 1/a)$ is determined from a global fit of zero-momentum matrix elements across multiple lattice spacings, relating the bare and renormalized matrix elements via a multiplicative factor:
\begin{equation} \label{eq:multaplicative_ZR}
    \tilde{h}^B (z, P_z=0, 1/a)
    =
    Z_R (z, 1/a) \tilde{h}^R (z, P_z=0, 1/a).
\end{equation}
where $Z_R$ is computed at one-loop order using the operator product expansion, with its precise form depending on the operator under consideration.
For our study, the self-renormalization factor takes the form
\begin{multline} \label{eq:renormalization-factor}
    Z_R (z, 1/a)
    =
    \text{exp} \biggl( \frac{kz}{a \ln [a\Lambda_{\text{QCD}}]} + m_0 z + f(z) a^2 \\
    + \frac{5 C_A}{3 b_0} \ln \left[ \frac{\ln [1/(a\Lambda_{\text{QCD}})]}{\ln [\mu / \Lambda_{\text{QCD}}]} \right] + \ln \left[ 1 + \frac{d}{\ln[a \Lambda_{\text{QCD}}]} \right] \biggr),
\end{multline}
where the first term is linearly divergent, the second term $m_0 z$ captures finite mass contributions from ambiguity in the renormalization, $f(z) a^2$ is dependent on the lattice action and takes into account discretization errors, and the remaining two terms originate from the resummation of leading and sub-leading logarithmic divergences.
In the $\overline{\text{MS}}$ scheme, the next-to-leading (NLO) order Wilson coefficient for the gluon can be derived from Ref.~\cite{Good:2025daz} as
\begin{equation} \label{eq:gluon-wilson-coeff}
    C_{0, \text{NLO}}(z, \mu)
    =
    1 + \frac{\alpha_s(\mu^2) C_A}{4 \pi} \left[ \frac{5}{3} \ln \left( \frac{z^2 \mu^2}{4 e^{-2 \gamma_E}} \right) + 3 \right],
\end{equation}
where $\gamma_E$ is the Euler-Mascheroni constant.
To determine the parameters in Eq.~\ref{eq:renormalization-factor}, we follow the approach of Ref.~\cite{Chen:2024rgi} and perform a global fit with the parameterization
\begin{equation}\label{eq:parameterization}
\begin{split}
    \ln \tilde{h}^B& (z, P_z = 0, 1/a)
    =
    \ln \left[ Z_R(z, 1/a)\right]  \\ &
    + \begin{cases}
       \ln[C_{0,\text{NLO}}(z,\mu)] , & z_0 \le z \le z_1, \\[2mm]
        g(z) - m_0 z, & z_1 < z.
    \end{cases}
\end{split}
\end{equation}
where we treat $k$, $\Lambda_\text{QCD}$, $f(z)$, $d$, $m_0$, and $g(z)$ as free fitting parameters.
We complete a global fit of these parameters to the interpolated zero momentum $\eta_s$ matrix elements for $a \approx  \{ 0.15, 0.12, 0.09 \}$~fm.
We use the $\eta_s$ data because it has better signal to noise than the nucleon, and the operator renormalization should not depend on the external state, and we confirmed that the sum of the linear divergence and renormalon ambiguity is statistically consistent between different hadrons in our previous LaMET study on only one lattice spacing~\cite{Good:2025daz}.
We choose $z_0 = 0.05$~fm and $z_1 = 0.36$~fm at a fixed renormalization scale $\mu = 2$~GeV, although scale dependence will be explored more carefully in the future.

\begin{figure}[t!]
\centering
\includegraphics[width=0.45\textwidth]{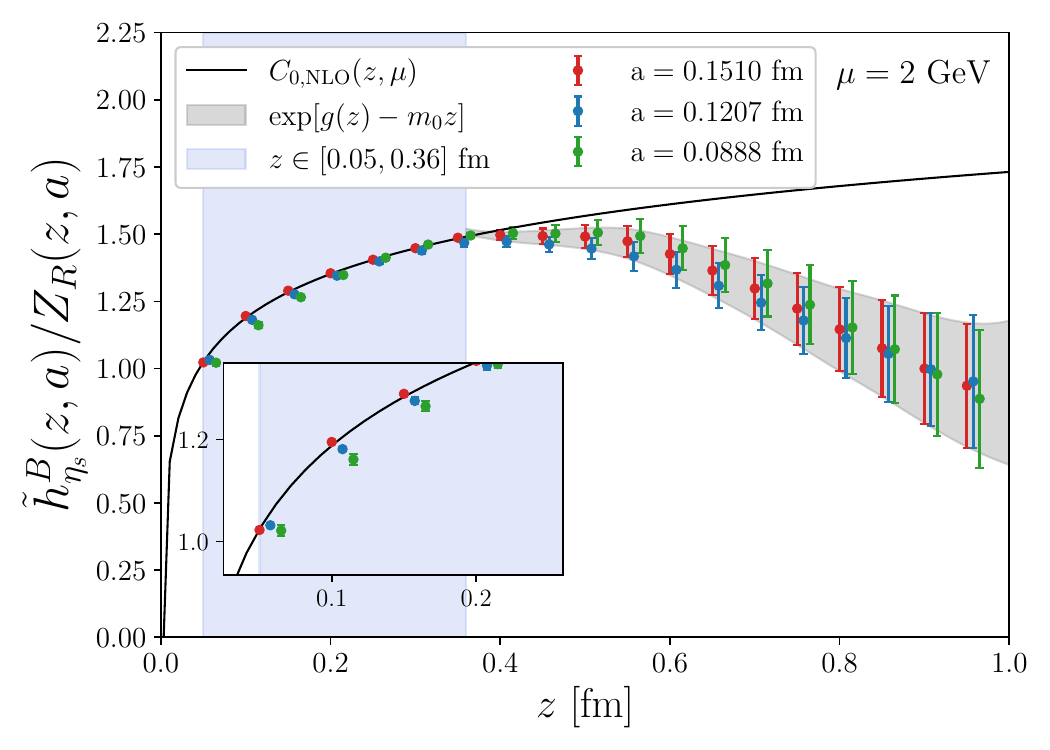}
\caption{
\label{fig:ZR_plots}
The renormalized matrix elements across each lattice spacing (data points) compared to the perturbative one-loop $\overline{\text{MS}}$ result illustrated by the black curve.
We denote the short distance region as $z \in [0.05, 0.36]$~fm as illustrated by the blue band, and the lattice-spacing--independent matrix element $\text{exp}[g(z) - m_0 z]$ by the grey band.
}
\end{figure}

The global-fit results in $k=0.61(27)$, $d = -0.002(10)$, $\Lambda_\text{QCD} = 0.286(85)$~GeV, and $m_0 = 0.13(30)$~GeV, with a $\chi^2/\text{dof} = 0.74(58)$, demonstrating that the fit is trustworthy.
All of the parameters fall within reasonable ranges as seen in previous quark calculations~\cite{LatticePartonLPC:2021gpi,Holligan:2024wpv,Chen:2024rgi, Holligan:2023rex}.
We find that $d$ is statistically consistent with zero likely due to this term being related to subleading logarithms, a fine-tuning parameter.
It should also be noted that the parameters are all correlated with each other, so the large standard deviations are not as suggestive of poor control over the fits as they may seem.
This can be seen in the original self-renormalization paper, which demonstrates that there are valleys of reasonable $\chi^2/\text{dof}$ for different values of $k$ and $\Lambda_\text{QCD}$, for example~\cite{LatticePartonLPC:2021gpi}.

In Fig.~\ref{fig:ZR_plots}, we show the renormalized zero momentum compared to the perturbative Wilson coefficient defined in Eq.~\ref{eq:gluon-wilson-coeff}.
We see that the matrix elements agree reasonably well with the perturbative Wilson coefficients in the short distance, while they begin to disagree in the long distance, both as one would expect.
Despite not being designed for smeared fields, the ability to apply self renormalization to this data is not so surprising.
In fact, under Wilson flow, the linear divergence is said to be proportional to  $1/\sqrt{\mathcal{T}_\text{W}}$~\cite{Brambilla:2023vwm}.
In the case of fixed relative flow time, $1/\sqrt{\mathcal{T}_\text{W}} \propto 1/a$, which can be absorbed into the first term for the linear divergence in Eq.~\ref{eq:renormalization-factor}.
Further more, if we also believe that the scale $a$ should be totally replaced by a scale $\propto\sqrt{\mathcal{T}_\text{W}}$ in the gradient flow scheme, then the entire self-renormalization expression in Eq.~\ref{eq:renormalization-factor}, would look nearly identical.
A more detailed and systematic study of this effect will be presented in forthcoming work~\cite{NieMiera:2025xxx}.

\textit{Renormalized matrix elements. ---}
We can now use the self-renormalization factor to renormalize the nucleon lattice data at finite momentum.
To obtain lightcone PDFs, we must first compute a Fourier transformation on the renormalized matrix elements to produce the quasi-PDF (defined in momentum space), then use the appropriate matching kernel to recover the lightcone PDF.
To avoid unphysical oscillations in the quasi-PDF, the renormalized matrix elements must be extrapolated to distances larger than those covered by our data.
We adopt the following extrapolation ansatz from Ref.~\cite{Ji:2020brr} which has strong theoretical motivations outlined in Ref.~\cite{Gao:2021dbh}:
\begin{equation}\label{eq:large-nu-form}
    h^\text{R}(z,P_z) \approx A\frac{e^{-m\nu}}{|\nu|^b}
\end{equation}
to describe the large-$\nu$ behavior of the data, where $A$, $m$, and $b$ are fit parameters.
We apply this fit form to the data from each ensemble for $z \gtrsim 0.6$~fm, finding $\chi^2/\text{dof} \approx 1$ for each fit.
To remove residual discretization and smearing artifacts, we then perform an extrapolation to the continuum limit $a \rightarrow 0$ using the ansatz
\begin{equation}
    \label{eq:cont_extrap}
    \tilde{h}^\text{latt}(z, P_z)
    =
    \tilde{h}^\text{cont}(z, P_z)
    + c(z) a^2,
\end{equation}
where $\tilde{h}^\text{cont}(z, P_z)$ denotes the continuum-extrapolated matrix element and $c(z)$ parametrizes the leading discretization effects, assumed to scale as $a^2$.
Additionally, we impose a positivity constraint on $\tilde{h}^\text{cont}(z, P_z)$ to ensure physical stability of the fit.
In our analysis, the large-$\nu$ extrapolations must be performed before the continuum extrapolation;
otherwise, the continuum matrix elements at large distances are too noisy to allow a reliable large-$\nu$ extrapolation.
These extrapolations should be commutative, but any associated systematics from this choice should be tested in the future with more precise data.

\begin{figure}
\centering
\includegraphics[width=0.475\textwidth]{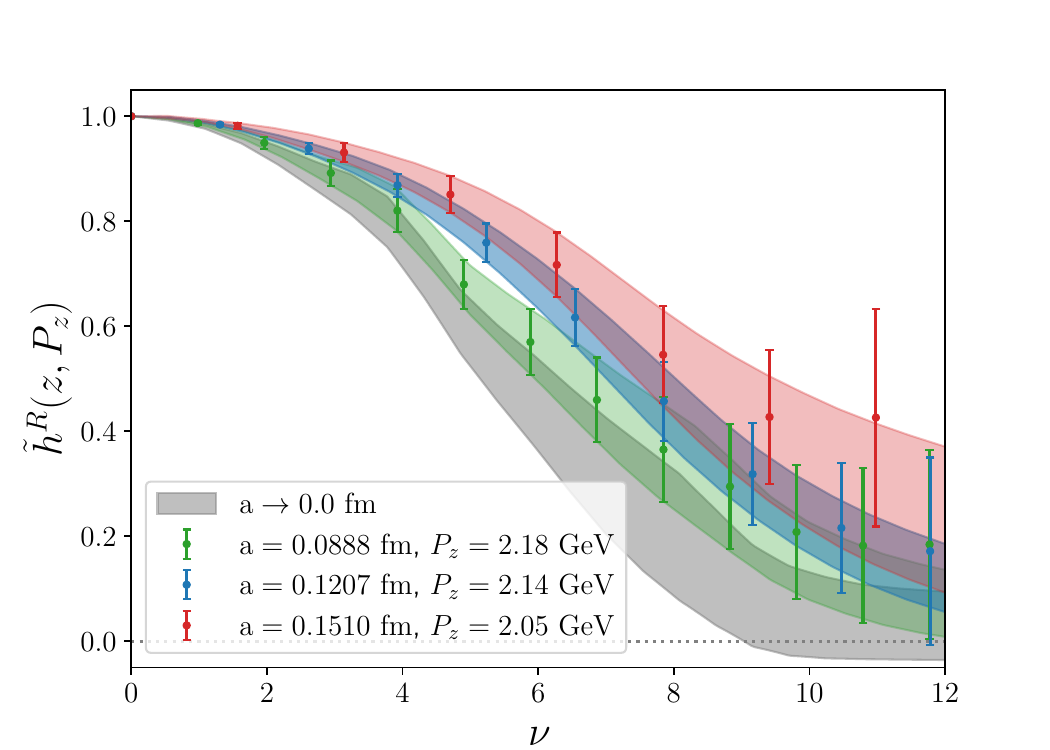}
\caption{
\label{fig:cont_extrap}
Renormalized lattice matrix elements (colored points) across three ensembles for $P_z \approx 2.0$-$2.2$~GeV compared to the continuous renormalized matrix elements (colored bands) obtained through the large-$\nu$ extrapolations and the continuum-extrapolated renormalized matrix element (gray band) obtained through Eq.~\ref{eq:cont_extrap}.
}
\end{figure}

In Fig.~\ref{fig:cont_extrap}, we present the discrete renormalized lattice data (colored points) for $P_z \approx 2.0$-$2.2$~GeV, compared with the corresponding continuous renormalized matrix elements (colored bands), which are obtained by combining the large-$\nu$ extrapolations with small-$\nu$ interpolations of the lattice data.
We see that the bands agree well with the lattice data and begin to exhibit exponential decay at large distances, though the rate of decay varies slightly between lattice spacings due to discretization and smearing effects.
The continuum-extrapolated matrix elements (gray band) follow the same trend, exhibiting increasingly rapid decay as the lattice spacing decreases.
Furthermore, the continuum results are in statistical agreement with the $a \approx 0.09$~fm matrix elements, indicating that the smallest lattice spacing is already nicely approaching the continuum limit, a trend that will be further constrained as we continue to reduce our lattice spacing in future work.

\begin{figure}
\includegraphics[width=0.45\textwidth]{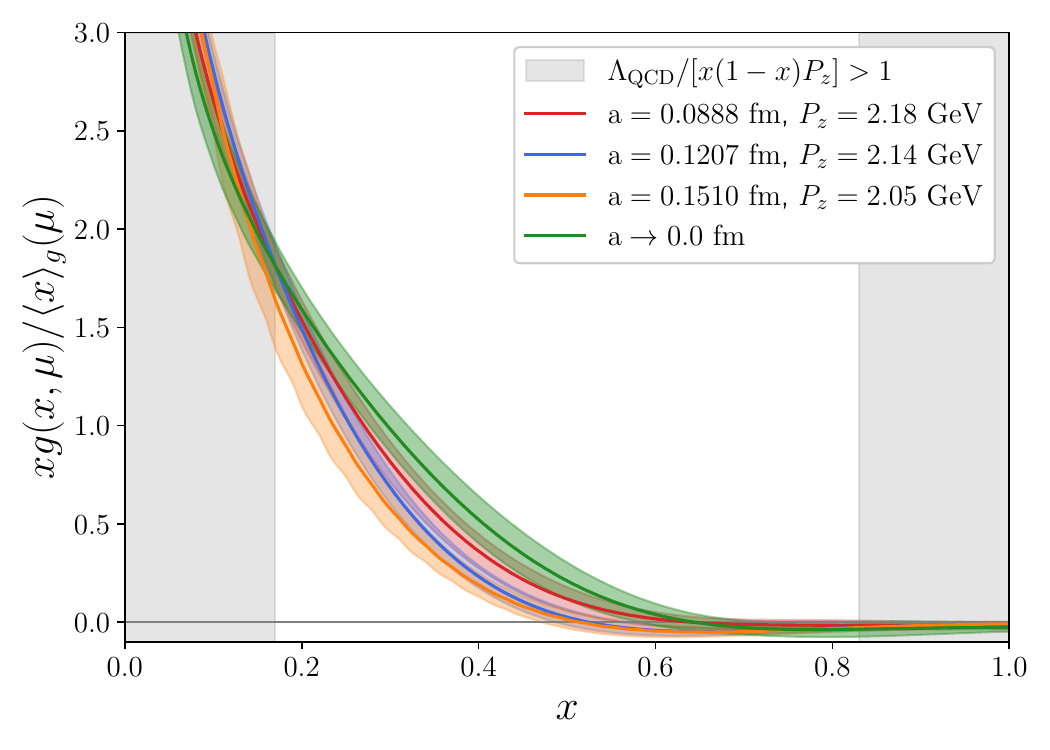}
\includegraphics[width=0.45\textwidth]{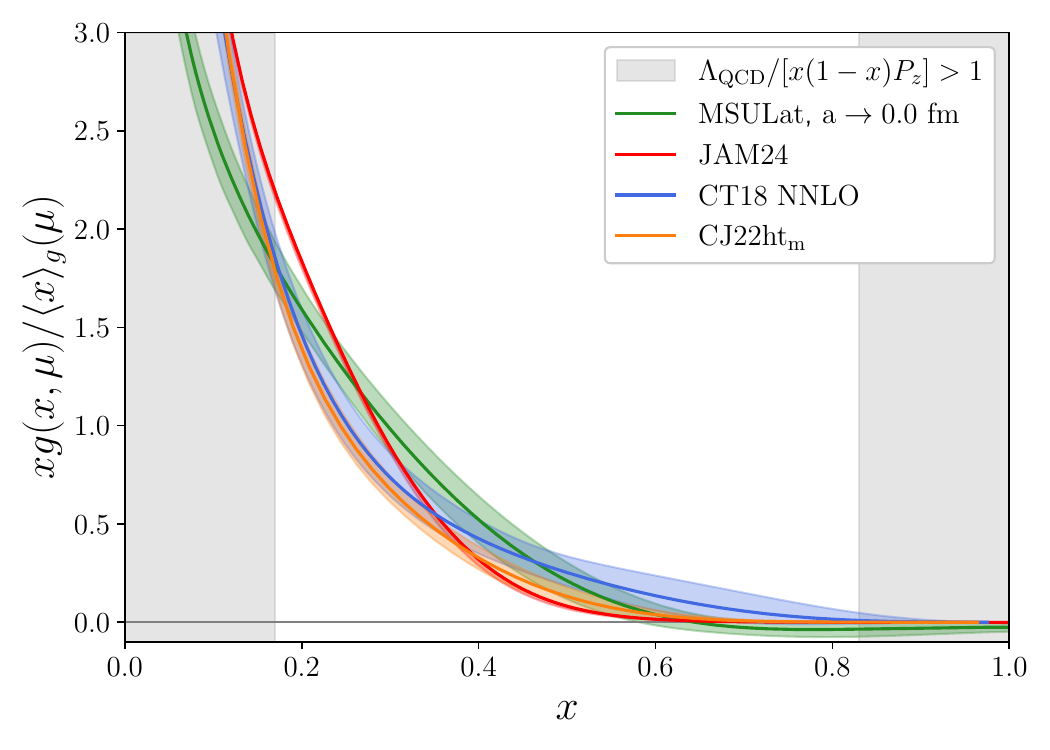}
\caption{
\label{fig:pheno_PDF_compare}
(upper) The MSULat unpolarized gluon PDF at $P_z = 2.12$~GeV in the continuum limit, shown alongside results obtained at individual lattice spacings.
(lower) The MSULat unpolarized gluon PDF at $P_z = 2.12$~GeV compared to select phenomenological determinations.
}
\end{figure}

\textit{Light-cone PDFs. ---}
With smooth, error-controlled, exponentially decaying data for each lattice spacing and the continuum, we can perform a Fourier transform to obtain the quasi-PDF.
These quasi-PDFs are then matched to the lightcone PDF using the procedure described in~\cite{Good:2025daz}, still neglecting quark mixing in this case.
For the continuum PDF, we perform the matching at $P_z = 2.12$~GeV, corresponding to the average of the three momenta, and the systematic uncertainty associated with these small ($< 10\%$) momentum differences is expected to be negligible.
The upper panel of Fig.~\ref{fig:pheno_PDF_compare} compares the PDFs from each lattice spacing with the continuum PDF, showing overall consistency with one another as expected from the underlying matrix elements.
For example, the $a\approx 0.15$~fm PDF (orange band) is roughly $1.2\sigma$ away from the the $a\approx 0.09$~fm PDF (green band).
The overall behavior of the PDFs across different lattice spacings naturally reflects the trends observed in the matrix elements.
In particular, since $x$ and $\nu$ are Fourier conjugates at leading order, slower decay of the matrix elements in $\nu$-space corresponds to a narrower distribution in $x$-space, and vice versa.
Thus, since the $a \approx 0.15$~fm matrix elements decay most slowly, the corresponding PDF is the narrowest, whereas the continuum-extrapolated PDF is the broadest.
In the lower panel of Fig.~\ref{fig:pheno_PDF_compare}, we compare our continuum-extrapolated gluon PDF to a selection of phenomenological gluon PDFs.
We consider the CT$18$~\cite{Hou:2019efy}, JAM$24$~\cite{Anderson:2024evk}, and the CJ$22$ multiplicative higher twist effect~\cite{Cerutti:2025yji} gluon PDFs as a representative sample of the global fits.
We note that our lattice PDF is slightly negative for $x \gtrsim 0.65$, but remains within $1\sigma$ of zero, and is expected to be controlled better when other systematics are considered.
We find that our lattice PDF is most consistent with the CT$18$ gluon PDF, with the difference between the two PDFs falling within 1$\sigma$ over the majority of the $x$-range where LaMET is trustworthy.
The lattice PDF falls as far as $2\sigma$-$3\sigma$ from the JAM$24$ and CJ$22$ PDFs in various regions below $x < 0.5$; however, it agrees with the very flat and near-zero behavior for $x > 0.6$, which implies that the density of gluons in the large-$x$ region for the proton is nearly zero.
This is in complement to recent work, supporting a higher density of large-$x$ gluons in the pion~\cite{Good:2025nny}
Assuming this overall behavior remains once additional systematics are explored and removed in our forthcoming work~\cite{NieMiera:2025xxx}, incorporating gluon PDFs from LaMET into phenomenological fits could provide significant constraints on the gluon PDF shape for $x > 0.2$.

\textit{Conclusions $\&$ Outlook. ---}
We have completed the first study of self-renormalization on the nucleon gluon PDF from LaMET in the continuum limit.
Interestingly, the self-renormalization method successfully absorbs the effects of Wilson flow at $\mathcal{T}_\text{W} = 3a^2$, despite not being originally designed with smearing in mind. 
This stability emphasizes the potential of self-renormalization and motivates further investigation into renormalization frameworks that explicitly account for smearing effects, which we plan to study in greater detail in forthcoming work~\cite{NieMiera:2025xxx}.
The renormalized matrix elements at $P_z \approx 2.0$-$2.2$~GeV across different lattice spacings are within $1\sigma$-$2\sigma$ of each other.
Thanks to the fixed relative flow time, the continuum extrapolation should simultaneously serve as a zero–flow time extrapolation, with the $a \approx 0.09$~fm data already statistically compatible with the zero–flow time continuum limit. 
Fourier transforming the renormalized matrix elements to obtain quasi-PDFs and matching them to the corresponding light-cone PDFs preserves this consistency, leading to agreement among the resulting PDFs across lattice spacing.
Finally, comparing our continuum-limit gluon PDF to the CT$18$~\cite{Hou:2019efy}, JAM$24$~\cite{Anderson:2024evk}, and CJ$22$~\cite{Cerutti:2025yji} global fits, we find that it exhibits a slightly flatter (less convex) shape in the region $0.2 < x < 0.5$, similar to CT$18$, and approaches near-zero values for $x > 0.6$, in agreement with JAM$24$ and CJ$22$.
These results are contingent on several systematics effects, including smearing, momentum, and pion-mass dependence, which we will explore in greater detail in a forthcoming companion paper~\cite{NieMiera:2025xxx}.

Looking ahead, our calculation can be further refined through improvements in lattice-QCD systematics within the LaMET framework, such as employing higher boost momenta, introducing additional nucleon operators, and utilizing ensembles with finer lattice spacings to better control the continuum extrapolation.
With enhanced signal-to-noise ratios for matrix elements in the large-$\nu$ region from upcoming calculations, it will become possible to explore alternative extrapolation formulations or apply Bayesian and machine-learning techniques~\cite{Chowdhury:2024ymm,Dutrieux:2024rem,Dutrieux:2025jed} to study their impact on the extracted PDFs.
Other systematic effects, such as leading-renormalon resummation~\cite{Zhang:2023bxs} and renormalization-group resummation~\cite{Su:2022fiu}, have been developed for the flavor-nonsinglet parton distributions (see the recent review in Ref.~\cite{Lin:2025hka} for example results), but are not yet available for the gluon PDF.
Extending these methods to the gluon sector would help suppress scale dependence and broaden the range of validity for the gluon PDF.
This work provides a major leap towards reaching a ``gold standard'' lattice calculation, which could significantly impact global fits of the gluon PDF.

\textit{Acknowledgments. ---}
AN and WG thank Yong Zhao, Yushan Su, Joshua Lin, and Jiunn-Wei Chen for their valuable suggestions and feedback on the earlier self-renormalized gluon PDF results.
We thank the MILC Collaboration for sharing the lattices used to perform this study.
The LQCD calculations were performed using the Chroma software suite~\cite{Edwards:2004sx}.
This research used resources of the National Energy Research Scientific Computing Center, a DOE Office of Science User Facility supported by the Office of Science of the U.S. Department of Energy under Contract No. DE-AC02-05CH11231 through ERCAP;
facilities of the USQCD Collaboration, which are funded by the Office of Science of the U.S. Department of Energy,
and supported in part by Michigan State University through computational resources provided by the Institute for Cyber-Enabled Research (iCER).
The work of AN and WG is partially supported by U.S. Department of Energy, Office of Science, under grant DE-SC0024053 ``High Energy Physics Computing Traineeship for Lattice Gauge Theory''.
The work of WG and HL is partially supported by the US National Science Foundation under grant PHY~2209424 and 2514533.
FY is supported by the U.S. Department of Energy, Office of Science, Office of Nuclear Physics through Contract No. DE-SC0012704, and within the framework of Scientific Discovery through Advanced Computing (SciDAC) award Fundamental Nuclear Physics at the Exascale and Beyond.

\bibliographystyle{unsrt}
\bibliography{main.bib}

\end{document}